\documentclass[12pt]{article}

\begin{document}

\title{Exact solutions for optimal execution of portfolios transactions  and the Riccati equation}

\author{ Juan M. Romero\thanks{jromero@correo.cua.uam.mx}  and Jorge Bautista\thanks{joba.89@hotmail.com} \\[0.5cm]
 \it Departamento de Matem\'aticas Aplicadas y Sistemas,\\
\it Universidad Aut\'onoma Metropolitana-Cuajimalpa,\\
\it M\'exico, D.F  05300, M\'exico\\
} 

\date{}

\pagestyle{plain}

\maketitle

\begin{abstract}
We propose  two   methods to obtain exact solutions for  the Almgren-Chriss model about optimal execution of portfolio transactions. In the first method we rewrite the  Almgren-Chriss equation and find two exact solutions. In the second method, employing a general  reparametrized time, we show that the 
 Almgren-Chriss equation can be reduced to some  known equations  which can be exactly solved in  different cases.
 For this last case we obtain a quantity conserved. In addition, we show that in both methods the Almgren-Chriss equation is equivalent to a  Riccati equation. 
 \end{abstract}

\section{Introduction}

In all financial phenomena there are many random variables, this is the reason why it is so difficult to construct mathematical models that provide realistic predictions in  markets. 
Frequently, when in a model all financial variables are considered, the model can not be tractable and then it is not useful. Meanwhile, when not all important variables are taken into account,  the model  may  not yield  realistic predictions.  Then, the challenge to construct a financial model is to consider  all important variables, but without sacrificing model tractability. It is worth mentioning that  in some cases it is possible to construct simple models  that provide 
realistic predictions. An example is given by the Black-Scholes equation for the european option price \cite{black,merton}.  Amazingly,   this equation can be exactly solved. Now, an important problem in finance is to obtain an optimal execution of portfolios transactions. In this regard, Merton proposed a model with constant volatility that can be analyticaly  solved  \cite{merton1,merton2}. The Merton's model has been extended for more realistic situations \cite{magill}, \cite{dumas}, \cite{liu}. 
A particular case of optimal execution problem is given  when an investor  wants to buy  (or sell) a large   amount of options with the  investment strategy  which provides  the  major profit.
In this case rapid buying  may increasing stock price while order  slicing  may add to  the uncertainty in the stoke  price. For this last problem, using the market impact  $\eta,$
the volatility $\sigma$  and supposing that the risk aversion $\lambda$ is a constant,   Almgren and Chriss showed that if  an investor at   time $t$   has  $x_{0}$ orders and at the final time $T$ must have  zero orders, the optimal cost for trading is given by the optimal value of the functional \cite{almgren1,almgren2}
 \begin{eqnarray}
 C=\int_{t}^{T}ds\left( \eta(s)\left(\frac{dx}{ds}\right)^{2}+\lambda \sigma^{2}(s) x^{2}(s)\right)  \label{eq:cost}.
 \end{eqnarray}
 From calculus of variations, it is well known  that a functional as $C$ reaches  its   optimal value when is evaluated in the function $x(s)$  which satisfies the Euler-Lagrange equation \cite{elsgoc}.   
For the cost (\ref{eq:cost}) the Euler-Lagrange is given by 
 \begin{eqnarray}
\eta(s)\frac{d^{2} x}{ds^{2}}+\frac{d\eta(s)}{ds}\frac{d x}{ds} =\lambda \sigma^{2}(s) x(s),
\label{eq:elal}
\end{eqnarray}
which must be solved with the boundary conditions 
 \begin{eqnarray}
 x(t)=x_{0}, \qquad x(T)=0.
\label{eq:elal-bc}
\end{eqnarray}
 The case when $\eta$ and $\sigma$ are constants can be exactly solved.
 However, when  $\eta$ and $\sigma$ are not  constants,  is not easy to find solutions to the equation (\ref{eq:elal}).
Now, it is worth pointing out that  some systems have a "natural time", for example the affine parameter to study geodesic curves \cite{eisen}. In fact, in some  financial models  the "natural time"  is  a stochastic time  \cite{gema}. 
 Frequently, the equation of motions are more friendly when are rewritten with  the "natural time". In this respect, Almgren and Chriss shown that  using  the time parameter 
\begin{eqnarray}
d\hat s=\sigma^{2}(s) ds  \label{eq:a-c-para}
\end{eqnarray}
and imposing the condition
\begin{eqnarray}
\eta(\hat s)\sigma^{2}(\hat s)={\rm constant}  \label{eq:a-c-c}
\end{eqnarray}
the  equation (\ref{eq:elal}) can be exactly  solved for some realistic cases. Some extensions of the  Almgren-Chriss model
can be seen in \cite{obizhaeva}, \cite{schied},\- \cite{gatheral1}, \\
\cite{gatheral2}.\\

 It is clear that to obtain optimal investment  strategies in the Almgren-Chriss model  is important to understand what kind of solutions  this  model  has. With this in mind, in this paper we propose  two   methods to obtain exact solutions for  the Almgren-Chriss equation. In the first method we rewrite the  Almgren-Chriss equation and find two exact solutions for this equation. Furthermore, 
using a Riccati equation,  we show that  the Almgren-Chriss can be solved. In this sense we can say that   the Almgren-Chriss equation is equivalent to a Riccati equation. In the second method we take the time as function of a general parameter, $\tau$, namely we take $s=s(\tau).$ Moreover, using a particular  time  parameter   we find an exact solution for  the Almgren-Chriss equation. Futhermore, we show that using a  "special  time"  the Almgren-Chriss equation  can be reduced a known equation which can be exactly solved in different cases. In addition we show that using this "special time" the Riccati equation equivalent to the Almgren-Chriss
equation can be  simplified. \\

This paper is organized as follow: in the section $2$ we provide the first method and obtain  two exact solutions for the 
Almgren-Chriss equation, in addition we show that the Almgren-Chriss equation is equivalent to a Riccati equation; 
 in the section $3$ we provide the second method  and obtain an exact solution for the Almgren-Chriss equation; in the section $4$ we give 
 a summary. 

 \section{First case }
 
 In order to obtain some exact solution for the Almgren-Chriss equation (\ref{eq:elal}) we propose 
 \begin{eqnarray}
x(s)= \frac{u(s)}{\sqrt{\eta(s)}}\label{eq:n-norm},
\end{eqnarray}
 in this case the cost (\ref{eq:cost}) becomes 
 \begin{eqnarray}
 C&=&\int_{t}^{T}ds \left( \left(\frac{du}{ds }\right)^{2}+\left( \frac{\frac{ d^{2} \eta(s)}{ds^{2}}}{2 \eta(s)} + \lambda \frac{\sigma^{2}(s)}{\eta(s) } 
 -\frac{\left(\frac{ d \eta(s)}{ds } \right) ^{2} }{4 \eta^{2}(s) }\right) u^{2}(s) \right)\nonumber\\
 &  & -\left( \frac{u^{2}(s)}{2} \frac{\frac{ d \eta(s)}{ds }  }{ \eta(s) }\right) \Bigg|_{t}^{T}. \label{eq:ac-nor}
 \end{eqnarray}
 From this expression   we obtain the following  Euler-Lagrange equation 
 \begin{eqnarray}
\frac{d^{2} u(s) }{ds ^{2}}=\left(\frac{\frac{ d^{2} \eta(s)}{ds^{2}}}{2 \eta(s)} + \lambda \frac{\sigma^{2}(s)}{\eta(s) } -\frac{\left(\frac{ d \eta(s)}{ds } \right) ^{2} }{4 \eta^{2}(s) }\right)  u(s ),\label{eq:norm}
\end{eqnarray}
which should be solved with the boundary condition
 \begin{eqnarray}
u(t)=x_{0}\sqrt{\eta(t)}, \qquad u(T)=0.
\label{eq:norm-bc}
\end{eqnarray}
 Now,  by an integration by parts, the cost (\ref{eq:ac-nor}) can be written as
 \begin{eqnarray}
 C&=&-\int_{t}^{T}ds  u(s)\left[\frac{d^{2}u}{ds^{2}} -\left( \frac{\frac{ d^{2} \eta(s)}{ds^{2}}}{2 \eta(s)} + \lambda \frac{\sigma^{2}(s)}{\eta(s) } -\frac{\left(\frac{ d \eta(s)}{ds } \right) ^{2} }{4 \eta^{2}(s) }\right) u(s)\right]\nonumber\\
 & &+ \left[u(s) \frac{du(s)}{ds} -\left( \frac{u^{2}(s)}{2} \frac{\frac{ d \eta(s)}{ds }  }{ \eta(s) }\right) \right]\Bigg|_{t}^{T}. 
 \end{eqnarray}
 Then,  if $u(s)$ satisfies the equation of motion (\ref{eq:norm}) and  the boundary conditions (\ref{eq:norm-bc}), we obtain 
 \begin{eqnarray}
 C&=&- \left[u(s) \frac{du(s)}{ds} -\left( \frac{u^{2}(s)}{2} \frac{\frac{ d \eta(s)}{ds }  }{ \eta(s) }\right) \right]\Bigg|_{s=t}.\label{eq:c-norm}
 \end{eqnarray}

 \subsection{Exact solution  1 }
 
 Let us consider the case 
 \begin{eqnarray}
 \eta(s)&=&\eta_{0}\gamma^{2} (\cosh as)^{2},\\
 \sigma(s)&=&\sigma_{0} \gamma \cosh as,
 \end{eqnarray}
 where $a, \eta_{0}, \gamma$ and  $\sigma_{0}$ are constants. Substituting  these functions in the equation (\ref{eq:norm}), 
 we get 
 \begin{eqnarray}
\frac{d^{2} u(s) }{ds ^{2}}=\left(a^{2}  +
\frac{ \lambda \sigma_{0}^{2}}{ \eta_{0} }  \right)  u(s ),
\end{eqnarray}
which is solved by the function  
 \begin{eqnarray}
u(s)=\frac{\sqrt{ \eta_{0}  } \gamma x_{0} (\cosh at )}{\sinh \sqrt{ a^{2}+\frac{ \lambda \sigma_{0}^{2} }{ \eta_{0} }} (T-t)  }
\sinh \sqrt{ a^{2}+\frac{\lambda \sigma_{0}^{2} }{ \eta_{0} }} (T- s).
 \end{eqnarray}
Furthermore, using the equation  (\ref{eq:n-norm}), we have 
 \begin{eqnarray}
x(s)= x_{0} \frac{  (\cosh at)}{  \sinh \sqrt{ a^{2}+\frac{ \lambda \sigma_{0}^{2} }{ \eta_{0} }} (T-t)  }\frac{ \sinh \sqrt{ a^{2}+\frac{\lambda \sigma_{0}^{2} }{ \eta_{0} }} (T-s) }{  \left( \cosh as \right)    } 
 \end{eqnarray}
and the cost (\ref{eq:c-norm}) is given by 
 \begin{eqnarray}
 C&=&\frac{\eta_{0} \gamma^{2} x_{0}^{2} \cosh^{2}(at) }{ \sinh \sqrt{ a^{2}+\frac{\lambda\sigma_{0}^{2} }{\eta_{0}}} (T-t) } \Bigg(  \sqrt{ a^{2}+\frac{\lambda\sigma_{0}^{2} }{\eta_{0}}} 
  \cosh \sqrt{ a^{2}+\frac{\lambda\sigma_{0}^{2} }{\eta_{0}}} (T-t)\nonumber\\
   & &+  a\left(  \frac{\sinh as}{\cosh as} \right)\sinh \sqrt{ a^{2}+\frac{\lambda\sigma_{0}^{2} }{\eta_{0}}} (T-t)  \Bigg)\nonumber\\
  &=&  \eta_{0} \gamma^{2} x_{0}^{2} \cosh^{2}(at) \left(  \sqrt{ a^{2}+\frac{\lambda\sigma_{0}^{2} }{\eta_{0}}} 
  \coth \sqrt{ a^{2}+\frac{\lambda\sigma_{0}^{2} }{\eta_{0}}} (T-t) +  a \tanh as\right) .\qquad 
  \end{eqnarray}

 \subsection{Exact solution  2}
 
Now, considering  the case 
 \begin{eqnarray}
\eta(s)&=&\eta_{0} e^{\zeta_{0} s},\\
\sigma(s)&=&\sigma_{0}e^{\frac{\zeta_{0} }{2} s}, \qquad \zeta_{0}={\rm constant},
\end{eqnarray}
 the equation (\ref{eq:norm}) becomes 
 \begin{eqnarray}
\frac{d^{2} u(s) }{ds ^{2}}=\left( \frac{\zeta^{2} _{0} }{4}+ \lambda \frac{\sigma^{2}_{0} }{\eta_{0} }\right)  u(s ),
\end{eqnarray}
which has the solution
 \begin{eqnarray}
u(s)=x_{0}\sqrt{\eta_{0} } e^{\frac{\zeta_{0}}{2} t} \frac{ \sinh \sqrt{ \frac{\zeta^{2}_{0}}{4}+\frac{\lambda \sigma_{0}^{2} }{ \eta_{0} }} (T-s)}{\sinh \sqrt{\frac{\zeta^{2}_{0}}{4}+\frac{ \lambda \sigma_{0}^{2} }{ \eta_{0} }} (T-t)  }.
 \end{eqnarray}
 In addition,  using the equation (\ref{eq:n-norm}),  we obtain 
 \begin{eqnarray}
x(s)=x_{0}e^{-\frac{\zeta_{0}}{2} (s-t) }\frac{ \sinh \sqrt{ \frac{\zeta^{2}_{0}}{4}+\frac{\lambda \sigma_{0}^{2} }{ \eta_{0} }} (T-s)}{\sinh \sqrt{\frac{\zeta^{2}_{0}}{4}+\frac{ \lambda \sigma_{0}^{2} }{ \eta_{0} }} (T-t)  }.
 \end{eqnarray}
Substituting this last  expression in the equation  (\ref{eq:c-norm}), we obtain the following cost 
 \begin{eqnarray}
 C&=&  \eta_{0}  x_{0}^{2} e^{\zeta_{0} t} \left(  \sqrt{ \frac{\zeta_{0}^{2}}{4}+\frac{\lambda\sigma_{0}^{2} }{\eta_{0}}} 
  \coth \sqrt{ \frac{\zeta_{0}^{2}}{4}+\frac{\lambda\sigma_{0}^{2} }{\eta_{0}}} (T-t) +   \frac{\zeta_{0}}{2}\right). \qquad 
  \end{eqnarray}

 \subsection{General  case and Ricatti equation}

 In general, given the functions $\sigma(s)$ and $\eta(s),$ to obtain a solution for the Almgren-Chriss equation 
 is a difficult task.  However, the problem  can be reduced to solve a known equation. In fact,  
 due that $\sigma$ and $\eta$ are positive functions, we can propose
 \begin{eqnarray}
\eta(s)=\eta_{0} e^{2\zeta_{1} (s)},\qquad \sigma(s)=\sigma_{0}e^{\zeta_{2} (s) }.
\end{eqnarray}
 Using these expressions in the equation (\ref{eq:norm}), we obtain 
 \begin{eqnarray}
\frac{d^{2} u(s) }{ds ^{2}}=\left( \frac{ d^{2} \zeta_{1} (s)}{ds^{2}}+ \left( \frac{ d \zeta_{1}(s)}{ds } \right) ^{2} +
\frac{ \lambda \sigma_{0}^{2}}{ \eta_{0} }  e^{2( \zeta_{2}(s)- \zeta_{1} (s))} \right)u(s). \label{eq:norm-ric}
\end{eqnarray}
This equation is solved by  
 \begin{eqnarray}
 u(s)=\frac{\sqrt{\eta(t)}  }{ e^{g(t)} \int_{t}^{ÊT} dz e^{-2g(z) } } e^{g(s)}\int_{s}^{ÊT} dz e^{-2g(z)},\label{eq:n-norm-s}
  \end{eqnarray}
where the function $g(s)$ must solve the equation
 \begin{eqnarray}
\frac{d^{2} g(s)}{ds^{2} }+\left( \frac{d g(s)}{ds }\right)^{2}=\left( \frac{ d^{2} \zeta_{1} (s)}{ds^{2}}+ \left( \frac{ d \zeta_{1}(s)}{ds } \right) ^{2} +
\frac{ \lambda \sigma_{0}^{2}}{ \eta_{0} }  e^{2( \zeta_{2}(s)- \zeta_{1} (s))} \right).\label{eq:r1}
  \end{eqnarray}
Notice that  in this case  the equation (\ref{eq:n-norm-s}) implies 
 \begin{eqnarray}
 x(s)=x_{0} \frac{ \sqrt{ \eta(t)}  } { e^{g(t)} \int_{t}^{ÊT} dz e^{-2g(z) }  }  \frac{ e^{g(s)} }{Ê\sqrt{\eta(s) }}   \int_{s}^{ÊT} dz e^{-2g(z)}.
 \end{eqnarray}
Furthermore,  if we take 
 \begin{eqnarray}
G(s)=\frac{d g(s)}{ds}
  \end{eqnarray}
the equation (\ref{eq:r1}) becomes 
 \begin{eqnarray}
\frac{d G(s)}{ds }+ G^{2}(s)=\omega^{2}(s) , \label{eq:r2}
  \end{eqnarray}
here 
 \begin{eqnarray}
\omega^{2}(s)=\left( \frac{ d^{2} \zeta_{1} (s)}{ds^{2}}+ \left( \frac{ d \zeta_{1}(s)}{ds } \right) ^{2} +
\frac{ \lambda \sigma_{0}^{2}}{ \eta_{0} }  e^{2( \zeta_{2}(s)- \zeta_{1} (s))} \right).
\end{eqnarray}
 We can see that the expression the expression  (\ref{eq:r2})  is a  Riccati equation \cite{arfken}.
Then, if the  Riccati  equation (\ref{eq:r2}) can be solved,  the equation (\ref{eq:norm-ric}) can be solved too. 
There is not a general method to solve the Riccati equation, however  some solutions for this equation are known.

 \section{Time reparametrization}
 
Some systems have a natural time, for instance  relaxation time in thermodynamic, mean lifetime in particle physicsand proper time in relativistic physics.  Frequently, using the natural time  the equations of motion are more kindly.\\

 In this section we propose a general time parameter and study different cases that can be exactly solved. 
 First, let us take the general parametrization  
 \begin{eqnarray}
s=s(\tau).
\end{eqnarray}
Using this parametrization, the cost (\ref{eq:cost}) becomes 
 \begin{eqnarray}
 C=\int_{\tau_{0}}^{\tau_{F}}d\tau \left( \frac{\eta(\tau)}{\frac{ds(\tau )}{d\tau} }\left(\frac{dx}{d\tau }\right)^{2}+\lambda \sigma^{2}(\tau)\frac{ds(\tau)}{d\tau} x^{2}(\tau) \right),\label{eq:crepa}
 \end{eqnarray}
which implies  the Euler-Lagrange equation
 \begin{eqnarray}
\frac{d^{2} x(\tau)}{d\tau^{2}}+\frac{dx(\tau)}{d\tau}\left( \frac{\frac{d\eta(\tau)}{d\tau }}{\eta(\tau)}-\frac{\frac{d^{2} s(\tau)}{d\tau^{2} }}{\frac{ds(\tau)}{d\tau}}\right)=\frac{ \lambda \sigma^{2}(\tau)}{\eta(\tau)}\left( \frac{ds(\tau)}{d\tau} \right)^{2} x(\tau).  \label{eq:repa}
 \end{eqnarray}
The boundary conditions for this last  equation are given by 
 \begin{eqnarray}
x(\tau_{0})=x_{0}, \qquad x(\tau_{F})=0. \label{eq:bc-repa}
 \end{eqnarray}
Notice that the cost (\ref{eq:crepa}) can be written as 
 \begin{eqnarray}
 C&=&-\int_{\tau_{0}}^{\tau_{F}}d\tau \frac{\eta(\tau) x(\tau)}{ \frac{ds(\tau)}{d\tau }} \Bigg[\frac{d^{2} x(\tau)}{d\tau^{2}}+\left( \frac{\frac{d\eta(\tau)}{d\tau }}{\eta(\tau)}-\frac{\frac{d^{2} s(\tau)}{d\tau^{2} }}{\frac{ds(\tau)}{d\tau}}\right)  \frac{dx(\tau)}{d\tau}  \nonumber\\
   & &- \frac{ \lambda \sigma^{2}(\tau)}{\eta(\tau)}\left( \frac{ds(\tau)}{d\tau} \right)^{2} x(\tau) \Bigg] 
  + \frac{\eta(\tau) x(\tau)}{ \frac{ds(\tau)}{d\tau }} \frac{dx(\tau)}{d\tau}\Bigg|_{\tau_{0}}^{\tau_{F} }.
  \end{eqnarray}
Then, when $x(\tau)$ satisfies the equation (\ref{eq:repa}) and the boundary conditions (\ref{eq:bc-repa}), we obtain the cost
 \begin{eqnarray}
 C&=&- \frac{\eta(\tau) x(\tau)}{ \frac{ds(\tau)}{d\tau }} \frac{dx(\tau)}{d\tau}\Bigg|_{\tau=\tau_{0}}.\label{eq:cos-rep}
  \end{eqnarray}

 We can see that taking the Almgren-Chriss parameter (\ref{eq:a-c-para}),
 that is 
 \begin{eqnarray}
 \frac{ds}{d\tau}=\frac{1}{\sigma^{2}(\tau)},
 \end{eqnarray}
 we obtain the cost 
 \begin{eqnarray}
 C=\int_{\tau_{0}}^{\tau_{F}}d\tau \left( \eta(\tau) \sigma^{2}(\tau)\left(\frac{dx}{d\tau }\right)^{2}+\lambda  x^{2}(\tau) \right),
 \end{eqnarray}
 and the Euler-Lagrange equation
 \begin{eqnarray}
\frac{d^{2} x(\tau)}{d\tau^{2}}+\frac{dx(\tau)}{d\tau}  \frac{d \ln\left( \eta(\tau) \sigma^{2}(\tau) \right)}{d\tau } 
=\frac{ \lambda}{\eta(\tau) \sigma^{2}(\tau)  } x(\tau).\label{eq:rep-ac}
 \end{eqnarray}
Almgren and Chriss shown that this equation is tractable when $\eta(\tau) \sigma^{2}(\tau)$ is a constant  
\cite{almgren1,almgren2}. \\
 
In the following subsections we introduce two additional useful time  parameters.

\subsection{First parameter } 
Due that $\sigma$ and $\eta$ are positive quantities, we can propose
\begin{eqnarray}
\eta(\tau)=\eta_{0}e^{\zeta_{1}(\tau) },\qquad \sigma(\tau)=\sigma_{0}e^{\zeta_{2}(\tau)}. \label{eq:im-vol}
\end{eqnarray}
 Then, the equation (\ref{eq:repa}) can be written as 
 \begin{eqnarray}
\frac{d^{2} x(\tau)}{d\tau^{2}}+\frac{dx(\tau)}{d\tau}\left( \frac{d\zeta_{1}(\tau)}{d\tau }-\frac{\frac{d^{2} s(\tau)}{d\tau^{2} }}{\frac{ds(\tau)}{d\tau}}\right)=\frac{ \lambda \sigma_{0}^{2}}{\eta_{0} }e^{ 2\zeta_{2}(\tau)-\zeta_{1}(\tau) }  \left( \frac{ds(\tau)}{d\tau} \right)^{2} x(\tau).
 \end{eqnarray}
Furthermore, using the parameter 
\begin{eqnarray}
\frac{ds(\tau)}{d\tau}=e^{\frac{   \zeta_{1}(\tau)-2\zeta_{2}(\tau)  }{2}},
\end{eqnarray}
we obtain
 \begin{eqnarray}
\frac{d^{2} x(\tau)}{d\tau^{2}}+\frac{dx(\tau)}{d\tau}\left( \frac{1}{2}\frac{d\zeta_{1}(\tau)}{d\tau }+\frac{d \zeta_{2} (\tau)}{d\tau }\right)=\frac{ \lambda \sigma_{0}^{2}}{\eta_{0} } x(\tau).\label{eq:repa1}
 \end{eqnarray}
 We can see  that when 
 \begin{eqnarray}
\frac{1}{2}\frac{d\zeta_{1}(\tau)}{d\tau }+\frac{d \zeta_{2} (\tau)}{d\tau }=\alpha, \qquad \alpha={\rm constant }
\end{eqnarray}
namely
 \begin{eqnarray}
\zeta_{2} (\tau)=\alpha \tau +\beta-\frac{\zeta_{1}(\tau)}{2}, \qquad \beta={\rm constant }, \label{eq:rep-con}
\end{eqnarray}
 the equation (\ref{eq:repa1}) becomes 
 \begin{eqnarray}
\frac{d^{2} x(\tau)}{d\tau^{2}}+\alpha \frac{dx(\tau)}{d\tau}=\frac{ \lambda \sigma_{0}^{2}}{\eta_{0} } x(\tau).\label{eq:expo}
 \end{eqnarray}
Notice that using the functions  (\ref{eq:im-vol}), the condition (\ref{eq:rep-con}) can be written as 
\begin{eqnarray}
\eta(\tau)\sigma^{2}(\tau)=A e^{2\alpha \tau},
\end{eqnarray}
 where $A$ is a constant. Furthermore, the solution for the equation (\ref{eq:expo}) is given by 
 \begin{eqnarray}
x(\tau)=x_{0}\frac{ e^{-\frac{\alpha}{2} (\tau-\tau_{0} )}\sinh\frac{1}{2} \sqrt{\alpha^{2}+ \frac{ 4\lambda \sigma_{0}^{2} }{ \eta_{0} }} (\tau_{F}-\tau)} {  \sinh\frac{1}{2} \sqrt{\alpha^{2}+ \frac{ 4\lambda \sigma_{0}^{2} }{ \eta_{0} }} (\tau_{F}-\tau_{0} )},
 \end{eqnarray}
while the cost is 
 \begin{eqnarray}
C=\frac{x_{0}^{2} }{2} \eta_{0} e^{\alpha \tau_{0} +\beta}\left( \alpha +\frac{1}{2} \sqrt{\alpha^{2}+ \frac{ 4\lambda \sigma_{0}^{2} }{ \eta_{0} }}  \coth\frac{1}{2} \sqrt{\alpha^{2}+ \frac{ 4\lambda \sigma_{0}^{2} }{ \eta_{0} }} (\tau_{F}-\tau_{0})\right) .
 \end{eqnarray}

 \subsection{Second parameter }	
 
Now, if we take the parameter 
 \begin{eqnarray}
\frac{ds}{d\tau}=\eta(\tau),
 \end{eqnarray}
the cost (\ref{eq:crepa}) becomes 
 \begin{eqnarray}
 C=\int_{\tau_{0}}^{\tau_{F}}d\tau \left( \left(\frac{dx}{d\tau }\right)^{2}+\lambda \sigma^{2}(\tau)\eta(\tau)  x^{2}(\tau) \right).
 \end{eqnarray}
This functional implies  the Euler-Lagrange equation 
 \begin{eqnarray}
\frac{d^{2} x(\tau) }{d\tau ^{2}}=\lambda \sigma^{2}(\tau) \eta(\tau) x(\tau ).\label{eq:aceq}
\end{eqnarray}
 Moreover,  when $x(\tau)$ satisfies the equation (\ref{eq:aceq}) and the boundary conditions (\ref{eq:bc-repa}), 
the cost  (\ref{eq:cos-rep}) becomes 
 \begin{eqnarray}
 C&=&- x(\tau) \frac{dx(\tau)}{d\tau}\Bigg|_{\tau=\tau_{0}}.
  \end{eqnarray}
 Notice that the original equation (\ref{eq:elal}) depends on  both functions $\sigma(s)$ and $\eta(s).$
 While the equation (\ref{eq:aceq}) only depends on the product  $\lambda \sigma^{2}(\tau) \eta(\tau).$ 
In addition, we can see that the equation (\ref{eq:aceq}) is simpler than the original equation (\ref{eq:elal}) and the 
equation (\ref{eq:rep-ac}). \\
 
 The equation (\ref{eq:aceq}) has some  interesting properties.   For example, if  the equation 
 \begin{eqnarray}
 \frac{ d^{2}\rho(\tau)}{d\tau^{2}  } +  \lambda \sigma^{2}(\tau) \eta(\tau) \rho(\tau)-\frac{1}{\rho^{3}(\tau)}=0
  \end{eqnarray}
 is satisfied, then   the function
 \begin{eqnarray}
 I= \frac{1}{2}\left[ \left(\rho(\tau) \frac{ dx(\tau)}{d\tau } -  x(\tau) \frac{ d\rho(\tau)}{d\tau }\right)^{2}- \left( \frac{ x(\tau)}{\rho (\tau)  }\right)^{2} \right],
   \end{eqnarray}
 is  a conserved quantity. In particular, when
 \begin{eqnarray} 
  \lambda \sigma^{2}(\tau) \eta(\tau)=\lambda \sigma_{0}^{2}\eta_{0}={\rm constant}, \label{eq:rep-con1}
  \end{eqnarray}
  we obtain the conserved quantity 
 \begin{eqnarray}
I= \frac{1}{2}\left[\frac{1}{\sqrt{\lambda \sigma_{0}^{2}\eta_{0} }} \left( \frac{ dx(\tau)}{d\tau }\right)^{2}- \sqrt{ \lambda \sigma_{0}^{2}\eta_{0}}x^{2}(\tau) \right],  
  \end{eqnarray}
which can be interpreted as  the energy for a particle with mass $ 1/ \sqrt{\lambda \sigma_{0}^{2}\eta_{0} }$ under the repulsive force $\sqrt{ \lambda \sigma_{0}^{2}\eta_{0}}x(\tau),$ see \cite{gol}. For the case (\ref{eq:rep-con1}),
we have 
 \begin{eqnarray}
x(\tau)= x_{0} \frac{\sinh \sqrt{\lambda\sigma_{0}^{2} \eta_{0}} ( \tau_{F} -\tau)   }{\sinh \sqrt{\lambda\sigma_{0}^{2} \eta_{0}} ( \tau_{F} -\tau_{0} ) }, 
  \end{eqnarray}
and the cost
 \begin{eqnarray}
C= x_{0}^{2} \sqrt{\lambda\sigma_{0}^{2} \eta_{0}}  \frac{\cosh  \sqrt{\lambda\sigma_{0}^{2} \eta_{0}} ( \tau_{F} -\tau_{0})   }{\sinh \sqrt{\lambda\sigma_{0}^{2} \eta_{0}} ( \tau_{F} -\tau_{0} ) }.
  \end{eqnarray}

In addition, it is possible to obtain solutions for  the equation  (\ref{eq:aceq}) for other  functions  $\lambda \sigma^{2}(\tau) \eta(\tau).$ 
 For instance, when 
 \begin{eqnarray}
\lambda \sigma^{2}(\tau) \eta(\tau)= \left(\alpha+\frac{\beta}{\tau +b}+\frac{\gamma }{(\tau+b)^{2} }\right) 
\end{eqnarray}
 where
 \begin{eqnarray}
\sqrt{1+4\gamma} \pm \frac{\beta }{\alpha }
\end{eqnarray}
is an integer, the equation  (\ref{eq:aceq}) is completely solved \cite{erma}.   Other  cases where the equation (\ref{eq:aceq}) is completely solved can be seen in \cite{erma}.\\

\subsection{Riccati equation}
Furthermore, in order to solve the equation (\ref{eq:aceq}), we can propose the function 
 \begin{eqnarray}
 x(\tau)=\frac{x_{0}}{ e^{f(\tau_{0} ) } \int_{\tau_{0}}^{Ê\tau_{F} } dz e^{-2f(z) } }  e^{f(\tau)} \int_{\tau}^{Ê\tau_{F} } dz e^{-2f(z)},
 \label{eq:sr}
  \end{eqnarray}
where the function $f(\tau)$ must satisfy  the equation
 \begin{eqnarray}
\frac{d^{2} f(\tau)}{d\tau^{2} }+\left( \frac{d f(\tau)}{d\tau }\right)^{2}=\lambda \sigma^{2}(\tau) \eta(\tau) .\label{eq:ac-r}
  \end{eqnarray}
 Thus, if we find a solution for this last equation, we can solve the equation (\ref{eq:aceq}). 
 In addition,  using the equation (\ref{eq:sr}) we obtain the cost 
 \begin{eqnarray}
C=x^{2}_{0} \left( \frac{  e^{-2 f(\tau_{0})} } {   \int_{\tau_{0}}^{\tau_{F} } dz e^{-2f(z) }  } -\frac{df(\tau)}{d \tau}\Bigg|_{\tau=\tau_{0}}  \right).
 \end{eqnarray}
 Notice that if we take 
 \begin{eqnarray}
F(\tau)=\frac{d f(\tau)}{d\tau}, 
  \end{eqnarray}
the equation (\ref{eq:ac-r}) can be written as 
 \begin{eqnarray}
\frac{d F(\tau)}{d\tau }+ F^{2}(\tau)=\lambda \sigma^{2}(\tau) \eta(\tau), \label{eq:rr2}
  \end{eqnarray}
which is a  Ricatti equation \cite{arfken}. Then, if the  Riccati  equation (\ref{eq:rr2}) can be solved,  the equation (\ref{eq:aceq}) can be solved too.\\

For instance, when 
 \begin{eqnarray}
\lambda \sigma^{2}(\tau) \eta(\tau)= \lambda \sigma^{2}_{0}  \eta_{0}\left( 1+\lambda \sigma^{2}_{0}  \eta_{0} \tau^{2}\right)
\end{eqnarray}
 we obtain 
 \begin{eqnarray}
F(\tau)=\lambda \sigma^{2}_{0}  \eta_{0} \tau,
\end{eqnarray}
 which implies 
 \begin{eqnarray}
f(\tau)=\lambda \sigma^{2}_{0}  \eta_{0} \frac{\tau^{2}}{2}+B,\qquad B={\rm constant}.
\end{eqnarray}
In this case, we have 
 \begin{eqnarray}
 x(\tau)=\frac{x_{0}}{  \int_{\tau_{0}}^{\tau_{F} } dz e^{- \lambda \sigma_{0}^{2} \eta_{0} z^{2} } } e^{\frac{ \lambda \sigma_{0}^{2}  \eta_{0} }{2} \left( \tau^{2}-\tau_{0}^{2}\right)  }  \int_{\tau}^{\tau_{F} } dz e^{-  \lambda \sigma_{0}^{2}  \eta_{0} z^{2} },
 \label{eq:sr}
  \end{eqnarray}
 and the cost is given by
 \begin{eqnarray}
C=x^{2}_{0} \left( \frac{  e^{- \lambda \sigma_{0}^{2} \eta_{0}  \tau_{0}^{2} } } {   \int_{\tau_{0}}^{\tau_{F} } dz e^{- \lambda \sigma_{0}^{2} \eta_{0}   z^{2}  }  } -  \lambda \sigma_{0}^{2} \eta_{0}  \tau_{0}  \right).
 \end{eqnarray}

 \section{Summary}
 In this paper  two   methods to obtain exact solutions for  the Almgren-Chriss equation were proposed. 
In the first method  the  Almgren-Chriss equation was rewritten and  two exact solutions for this equation were found. Furthermore, 
using a Riccati equation,  we shown that  the Almgren-Chriss can be solved. In this sense we can say that   the Almgren-Chriss equation is equivalent to a Riccati equation. In the second method the Almgren-Chriss equation was reparametrized on the time.   Moreover, using a particular parameter "time"  we found an exact solution for  the Almgren-Chriss equation. Futhermore, we shown that using a  special  reparametrization   the Almgren-Chriss equation  can be reduced to a known equation which can be exactly solved for different cases. For this last case we  obtained a 
conserved quantity.

\section*{Acknowledgments}

This work was supported in part by  Conacyt-SEP project  CB-2011-180111 (J.M.R). 
%


\end{document}